\documentclass[fleqn,usenatbib]{mnras}

\usepackage{newtxtext}
\usepackage{xcolor}
\usepackage{graphicx}

\usepackage[T1]{fontenc}

\DeclareRobustCommand{\VAN}[3]{#2}
\let\VANthebibliography\thebibliography
\def\thebibliography{\DeclareRobustCommand{\VAN}[3]{##3}\VANthebibliography}

\usepackage{graphicx}	
\usepackage{amsmath}	
\usepackage{amssymb}	

\newcommand{\angstrom}{{\rm \AA}}

\newcommand{\CIV}{C\,{\small IV}\,$\lambda$1549}

\newcommand{\CIII}{C\,{\small III}]\,$\lambda$1909}
\newcommand{\MgII}{Mg\,{\small II}\,$\lambda$2800}

\newcommand{\hbeta}{H{$\beta$}}

\title[Fermi Blazar Mass]{Gemini Near-infrared Spectroscopy of High-Redshift Fermi Blazars: Jetted Black Holes in the Early Universe Were Overly Massive}

\author[Burke et al.]{
Colin J. Burke,$^{1}$
Xin Liu,$^{1,2,3}$\thanks{E-mail: xinliuxl@illinois.edu}
Yue Shen$^{1,2}$
\\
$^{1}$Department of Astronomy, University of Illinois at Urbana-Champaign, 1002 West Green Street, Urbana, IL 61801, USA\\
$^{2}$National Center for Supercomputing Applications, 1205 West Clark Street, Urbana, IL 61801, USA\\
$^{3}$Center for Artificial Intelligence Innovation, University of Illinois at Urbana-Champaign, 1205 West Clark Street, Urbana, IL 61801, USA \\
}

\date{Accepted XXX. Received YYY; in original form ZZZ}

\pubyear{2023}

\begin{document}
\label{firstpage}
\pagerange{\pageref{firstpage}--\pageref{lastpage}}
\maketitle

\begin{abstract}
Jetted active galactic nuclei (AGNs) are the principal extragalactic $\gamma$-ray sources. Fermi-detected high-redshift ($z>3$) blazars are jetted AGNs thought to be powered by massive, rapidly spinning supermassive black holes (SMBHs) in the early universe ($<2$ Gyr). They provide a laboratory to study early black hole (BH) growth and super-Eddington accretion -- possibly responsible for the more rapid formation of jetted BHs. However, previous virial BH masses of $z>3$ blazars were based on \CIV\ in the observed optical, but \CIV\ is known to be biased by strong outflows. We present new Gemini/GNIRS near-IR spectroscopy for a sample of nine $z>3$ Fermi $\gamma$-ray blazars with available multi-wavelength observations that maximally sample the spectral energy distributions (SEDs). We estimate virial BH masses based on the better calibrated broad H$\beta$ and/or \MgII . We compare the new virial BH masses against independent mass estimates from SED modeling. Our work represents the first step in campaigning for more robust virial BH masses and Eddington ratios for high-redshift Fermi blazars. Our new results confirm that high-redshift Fermi blazars indeed host overly massive SMBHs as suggested by previous work, which may pose a theoretical challenge for models of the rapid early growth of jetted SMBHs.
\end{abstract}




\section{Introduction} \label{sec:intro}

Active supermassive black holes (SMBHs) with masses $\gtrsim10^9 M_{\odot}$ powering the most luminous quasars have formed when the universe was less than a Gyr after the Big Bang (e.g., \citealt{Wang2021}). Understanding how SMBHs formed so quickly is a major outstanding problem in modern astrophysics \citep{Natarajan2014,Reines2016,Inayoshi2020}, because their presence in large number densities may pose a challenge to modeling their rapid formation and subsequent growth \citep{Johnson2016}. High-redshift ($z>3$) blazars are ideal laboratories to study early SMBH growth within the first 2 Gyr of the universe. These sources are active galactic nuclei (AGNs) powered by billion solar mass black holes with our line of sight lying within an angle $\sim1/\Gamma$ of the jet axis, where $\Gamma$ is the jet bulk Lorentz factor ($\Gamma \approx 10$--15; \citealt{Ghisellini2014a}).

The Fermi Large Area Telescope (Fermi-LAT; \citealt{Atwood2009}) has detected high-redshift ($z>3$) blazars, which are dominated by flat-spectrum radio quasars \citep{Ackermann2017} with measured BH masses $\gtrsim 10^9 M_\odot$ \citep{Marcotulli2020}. The observation of a single blazar implies the presence of $\sim2 \Gamma^2$ misaligned, jetted systems with the same BH mass pointing at other directions, increasing the space density of SMBHs of jetted AGN by 10-20 percent depending on the redshift bin \citep{Sbarrato2015,Ackermann2017}. The space density of high-redshift Fermi blazars implies that radio-loud systems with SMBHs $\gtrsim 10^9 M_{\odot}$ are as common as radio-quiet systems, and they may be even more common at higher redshifts \citep{Sbarrato2015}. Therefore, the jetted phase is likely a key ingredient for understanding rapid black hole growth in the early universe. 

While high-redshift Fermi blazars are not as distant as the highest-redshift quasars known, they still pose a challenge for early SMBH growth models. Jetted BHs are thought to be less efficient accretors due to higher radiative loss (e.g., from having a larger spin, if jet activity is correlated with the BH spin). For example, a spinning BH accreting at Eddington rate would need $\sim$3.1 Gyr to grow from a seed of $\sim100 \,M_{\odot}$ to $\sim10^9 \,M_{\odot}$ \citep{Ghisellini2013}. This implies that such massive, jetted BHs would not have had formed at $z>2$, whereas their observed space density peaks around $z\sim4$ \citep{Sbarrato2015}. Simulations suggest that BHs accreting at super-Eddington rates are characterized by strongly collimated outflows or jets \citep{McKinney2014,Scadowski2014}. High-redshift blazars provide a laboratory to study super-Eddington accretion which may have led to more rapid growth of the jetted SMBHs than non-jetted systems \citep{Volonteri2015}.

Accurate BH mass measurements are crucial in order to quantify blazar demographics and growth. However, the existing BH mass estimates of high-redshift blazars are highly uncertain. As listed in Table~\ref{tab:prop1} and ~\ref{tab:prop2}, the existing BH masses of these high-redshift blazars were estimated using two methods: SED modeling of the accretion disk, and virial mass using \CIV\ \citep{Paliya2020}. Both estimates may have significant problems. The SED masses are limited by: (1) the data quality to cleanly (with jet contamination) and fully (limited by the Ly$\alpha$ forest absorption at the higher frequency range) sample the accretion disk emission and (2) the assumption of the standard disk model \citep{Shakura1973}, which may break down for highly spinning BHs. 

We present new Gemini/GNIRS NIR spectroscopy to obtain robust virial BH masses for nine Fermi blazars known at $z>3$ with existing multi-wavelength observations that maximally sample the SEDs \citep{Paliya2020}. For the virial BH mass, \CIV\ is the only broad emission line covered by the existing optical spectra for $z>3$ blazars \citep{Ackermann2017}. However, it is well known that \CIV\ is a poor virial mass estimator \citep{Shen2012,Jun2015a}. While a general agreement is found between the SED-based and virial BH masses in 116 Fermi blazars at $z<3.2$ \citep{Ghisellini2015a}, the virial masses of $>95$\% of the sample are based on H$\beta$ and/or \MgII , which do not suffer from the biases of \CIV . Indeed, the few objects with \CIV -based masses in low-redshift blazars exhibit the largest discrepancy with the SED-based masses.  \CIV\ often shows significant blueshifts due to accretion disk winds and strong outflows, which may be particularly relevant for high-redshift blazars. Significant mass uncertainties induce errors in the mass function and Eddington ratio distribution, hampering a robust test of theories of early SMBH formation and growth. Therefore, we will use Gemini/GNIRS NIR spectroscopy to measure the broad H$\beta$ and \MgII\ to obtain robust virial BH masses. The sample selection, observations, and data analysis are presented in \S\ref{sec:data}, our resulting virial BH masses, bolometric luminosities, and Eddington ratios, along with comparison with the SDSS spectra and SED fitting are presented in \S\ref{sec:results}, and we conclude in \S\ref{sec:conclusion}.

\section{Observations and Data Analysis} \label{sec:data}

\subsection{Sample Selection}

The targets are the farthest known Fermi-detected $\gamma$-ray blazars. A $\gamma$-ray detection is a definitive signature for the presence of a closely aligned relativistic jet. The targets are the nine $\gamma$-ray detected blazars presented in \citep{Paliya2020}. They were identified in a systematic search of $\gamma$-ray counterparts detected by the Large Area Telescope on board Fermi \citep{Atwood2009} for all the $z>3$ radio-loud ($R>10$, where $R$ is the ratio of the rest-frame 5 GHz to optical $B$-band flux density) quasars in the Million Quasar Catalog \citep{Flesch2015}. Compared with the Fermi blazars at lower redshifts, the $z>3$ Fermi blazars occupy the region of high $\gamma$-ray luminosities and soft photon indices, typical of powerful blazars; their Compton dominance (i.e., ratio of the inverse Compton to synchrotron peak luminosities) is large ($>20$), placing them among the most extreme blazar population \citep{Ackermann2017}. 

The targets are all flat spectrum radio quasars with prominent emission lines \citep{urry95} (rest-frame EW$>5$\angstrom ) suitable for virial BH mass measurements. They also have existing X-ray observations from the Chandra, XMM-Newton, and/or Swift-XRT data archives for maximally sampling the multi-wavelength SEDs. The targets include nine of the 10 $\gamma$-ray detected $z>3$ blazars in the fourth catalog of the Fermi-LAT-detected AGNs \citep{Ajello2020}. The remaining object is not included because it lacks existing X-ray data. While there are X-ray selected blazars at even higher redshifts, they are less suitable for this program because their SEDs are not as fully sampled. 

\begin{table*}
\caption{{\small Source properties for our blazar sample with GNIRS spectra. The SED-based masses (typical uncertainty of 0.3 dex) and accretion disk luminosities (typical uncertainty of 0.4 dex) are from \citet{Paliya2020}. The continuum luminosities are measured from our GNIRS spectra at 5100 \AA\ or $^{1}$ 3000 \AA\ otherwise. The uncertainties on the continuum luminosities are measurement errors only. True uncertainties are dominated by systematic differences in data reduction and spectral fitting choices. Virial BH masses are given in Table~\ref{tab:prop2}.}}
\centering
\begin{tabular}{lcccccccc}
\hline\hline
 & $z$ & $M_{{\rm BH,\ SED}}$ & $L_{\rm{disk}}$ & $R$ & $L_{5100}$ & S/N$_{{\rm C~IV}}$ & S/N$_{{\rm Mg~II}}$ & S/N$_{{\rm H}\beta}$  \\
Target & & (log$M_{\odot}$) & (erg s$^{-1}$) & (mag) & (erg s$^{-1}$) &  &  \\ 
\hline
J0337$-$1204 & 3.44 & 9.0 & 46.36 & 20.19 & $46.050{\pm}0.005$         &  --  & 0.8 & 1.2 \\
J0539$-$2839 & 3.10 & 9.3 & 46.70 & 18.97 & $46.828{\pm}0.001$         &  --  & 5.5 & 0.7 \\
J0733$+$0456 & 3.01 & 8.7 & 46.60 & 18.76 & $46.038{\pm}0.002$         & --  & 3.3 & 2.7 \\
J0805$+$6144 & 3.03 & 9.0 & 46.34 & 19.81 & $46.186{\pm}0.002$         & --  & 2.6 & 1.3 \\
J0833$-$0454 & 3.45 & 9.5 & 47.15 & 18.68 & $46.517{\pm}0.002$         &  --  & 6.9 & 4.4 \\
J1354$-$0206 & 3.72 & 9.0 & 46.78 & 19.64 & $46.192{\pm}0.001$         & 0.7 & 0.4 & 1.4 \\
J1429$+$5406 & 3.03 & 9.0 & 46.26 & 19.84 & $45.935{\pm}0.003$         & 2.0 & 1.3 & 2.2 \\
J1510$+$5702 & 4.31 & 9.6 & 46.63 & 19.89 & $\ \ 46.621{\pm}0.003^{1}$ & 2.4 & 0.6 & --  \\
J1635$+$3629 & 3.62 & 9.5 & 46.30 & 20.55 & $46.144{\pm}0.001$         & 2.6 & 4.9 & 0.8 \\
\hline
\end{tabular}
\label{tab:prop1}
\end{table*}

\begin{table*}
\caption{{\small Virial BH masses measured from broad-line detections (line $\rm{S/N} > 2$) in our GNIRS (Mg~II, H$\beta$) or SDSS (C~IV) spectra. All uncertainties are measurement errors only. True uncertainties are dominated by systematic differences in data reduction and spectral fitting choices. Missing table entries either do not have sufficient line detections to estimate a BH mass or are not covered within the wavelength range of the spectrum.}}
\centering
\begin{tabular}{lccccccccccc}
\hline\hline
& $M_{{\rm BH,\ C~IV,\ cont}}$ & $M_{{\rm BH,\ Mg~II,\ cont}}$ & $M_{{\rm BH,\ H}\beta{\rm,\ cont}}$ & $M_{{\rm BH,\ C~IV,\ line}}$ & $M_{{\rm BH,\ Mg~II,\ line}}$ & $M_{{\rm BH,\ H}\beta{\rm,\ line}}$  \\
Target & (log$M_{\odot}$) & (log$M_{\odot}$) & (log$M_{\odot}$) & (log$M_{\odot}$) & (log$M_{\odot}$) & (log$M_{\odot}$) \\ 
\hline
J0337$-$1204 & --              & --              & --              & --              & --              & -- \\
J0539$-$2839 & --              & $9.60{\pm}0.10$ & --              & --              & $9.01{\pm}0.12$ & -- \\
J0733$+$0456 & --              & $8.90{\pm}0.04$ & $9.32{\pm}0.25$ & --              & $8.64{\pm}0.05$ & $9.02{\pm}0.28$ \\
J0805$+$6144 & --              & $9.71{\pm}0.14$ & --              & --              & $9.48{\pm}0.15$ & -- \\
J0833$-$0454 & --              & $9.43{\pm}0.03$ & $9.27{\pm}0.06$ & --              & $9.22{\pm}0.04$ & $8.91{\pm}0.07$ \\
J1354$-$0206 & --              & --              & --              & --              & --              & -- \\
J1429$+$5406 & $8.91{\pm}0.81$ & --              & $8.85{\pm}0.15$ & $8.85{\pm}0.82$ & --              & $8.53{\pm}0.18$ \\
J1510$+$5702 & $9.54{\pm}0.29$ & --              & --              & $9.44{\pm}0.28$ & --              & -- \\
J1635$+$3629 & $9.12{\pm}0.10$ & $9.93{\pm}0.07$ & --              & $9.05{\pm}0.11$ & $9.82{\pm}0.07$ & -- \\
\hline
\end{tabular}
\label{tab:prop2}
\end{table*}

\subsection{SDSS Spectroscopy}

We searched the SDSS data release 18 database for optical spectra. Four of our blazars, J1354$-$0206, J1429$+$5406, J1510$+$5702, and J1635$+$3629 have SDSS spectra, from which we will obtain \CIV\ BH mass measurements.

\subsection{Gemini/GNIRS Observations}

We conducted Gemini-N/GNIRS NIR spectroscopy for the nine $z>3$ Fermi blazars with the goal to measure H$\beta$ (\MgII ) in the $K$ ($J$) band for the eight targets at redshift $3.0<z<3.7$ and \MgII\ in the $H$ band for the $z=4.31$ target in regions of high atmospheric transparency. We used the 32 line mm$^{-1}$ grating with a cross-dispersion and a 0.$''$675 slit. This gives a spectral resolution of $R\approx800$ and covers the full J, H, and K bands in the observed 0.8--2.5 $\mu$m using nodded exposures along the slit for 120 s each. The observing conditions and resulting data quality varied. We were unable to detect broad emission lines in the NIR for three sources: J0337-1204, J1354-0206, and J1510+5702.




\subsection{Data Reduction and Analysis}

\begin{figure*}
\includegraphics[width=0.98\textwidth]{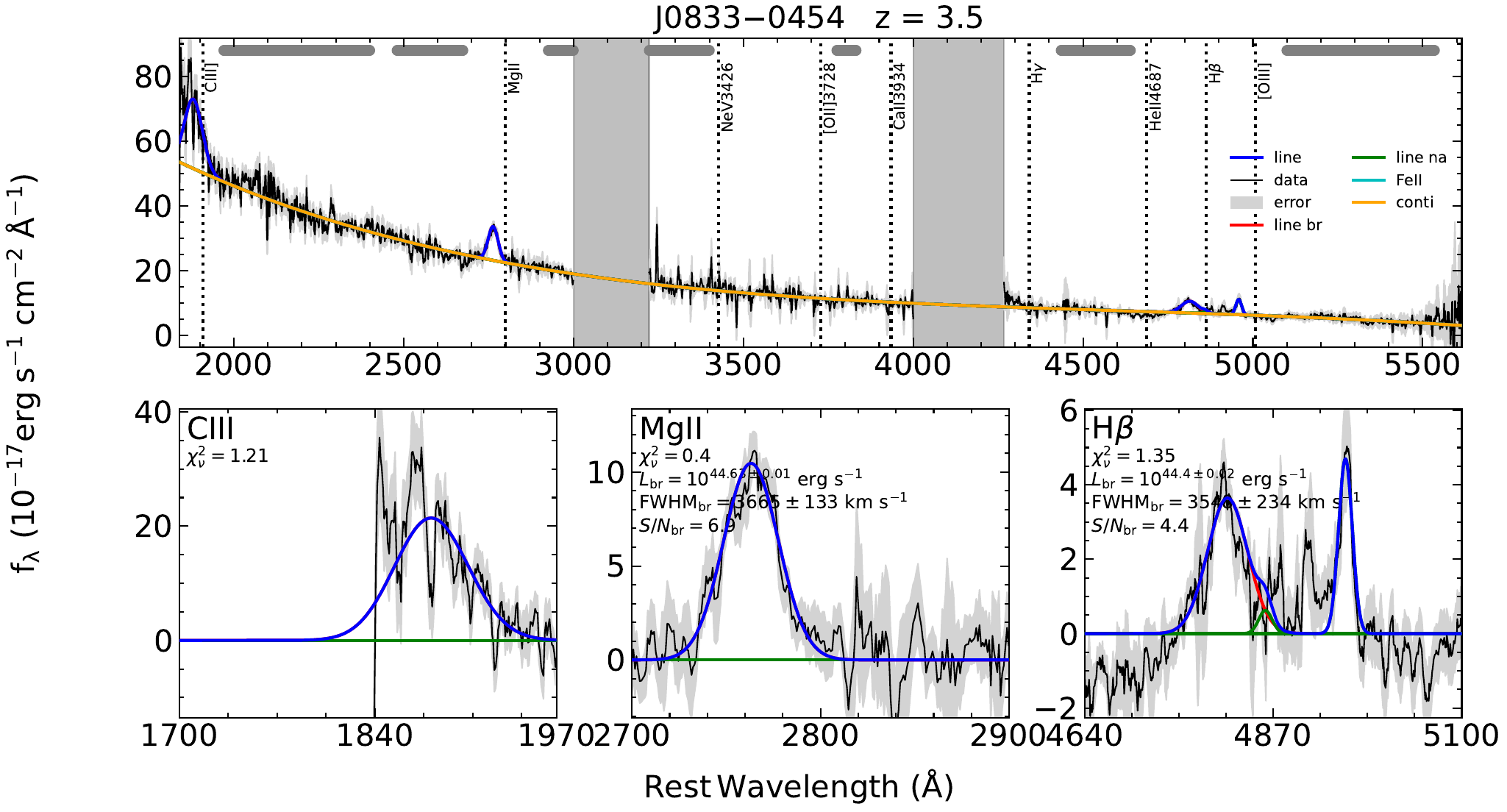}
\caption{Example GNIRS spectrum of the source J0833$-$0454, the brightest in our sample. A power-law plus 3rd-order polynomial and Gaussians are used to fit the continuum (yellow) and emission lines, respectively. The Fe II emission templates (teal) are set to zero, as including them does not significantly improve the fits. The data is shown in black and the best-fit model is overplotted in blue. The individual narrow line components are plotted in green, and the broad line components are plotted in red. The gray bands on the top are line-free windows selected to determine the continuum emission. The light gray shaded bands are masked regions effected by telluric absorption at 1.35$-$1.45 $\mu$m and 1.8$-$1.92 $\mu$m. The lower panels show the zoomed-in emission line regions of \CIII , \MgII , and \hbeta . The full figure set showing the fitting results for each source in our sample is shown in Appendix~\ref{app:nir}.}
\label{fig:spec}
\end{figure*}

We used the \textsc{pypeit} version 1.10.0 package \citep{pypeit:zenodo,pypeit:joss_arXiv,pypeit:joss_pub} to reduce the GNIRS spectra. The reduction pipeline steps include sky subtraction using standard A$-$B mode subtraction and a B-spline fitting, wavelength calibration using night-sky OH lines, cosmic ray rejection with \textsc{l.a.~cosmic} \citep{vanDokkum2012}, flux calibration using spectroscopic standard stars, and telluric correction derived from fitting telluric models from the Line-By-Line Radiative Transfer Model (\textsc{lblrtm}; \citealt{Clough2005}). The 1D spectra are extracted following the method of \citet{Horne1986}.

We fit the continuum and emission lines in each 1D spectrum using a modified version of the publicly-available \textsc{PyQSOFit} code \citep{Guo2018,Shen2019}. In this code, the continuum is modeled as a blue power-law plus a 3rd-order polynomial for reddening. No Fe~II continuum emission was detected, and including it does not significantly improve the fits, so Fe~II emission templates \citep{Vestergaard2001} were not used. The total model is a linear combination of the continuum and single or multiple Gaussians for the emission lines. Since uncertainties in the continuum model may induce subtle effects on measurements for weak emission lines, we first perform a global fit to the emission-line free region to better quantify the continuum. We then fit multiple Gaussian models to the continuum-subtracted spectra around the \hbeta\ and \MgII\ emission line complex regions locally. We define narrow Gaussians as having FWHM $< 1800$ km s$^{-1}$. The narrow line widths are tied together. We use 50 Monte Carlo simulations to estimate the uncertainty in the line measurements. Figure~\ref{fig:spec} shows an example GNIRS spectra of J0833$-$0454, the brightest source in our sample.

\begin{figure}
\includegraphics[width=0.48\textwidth]{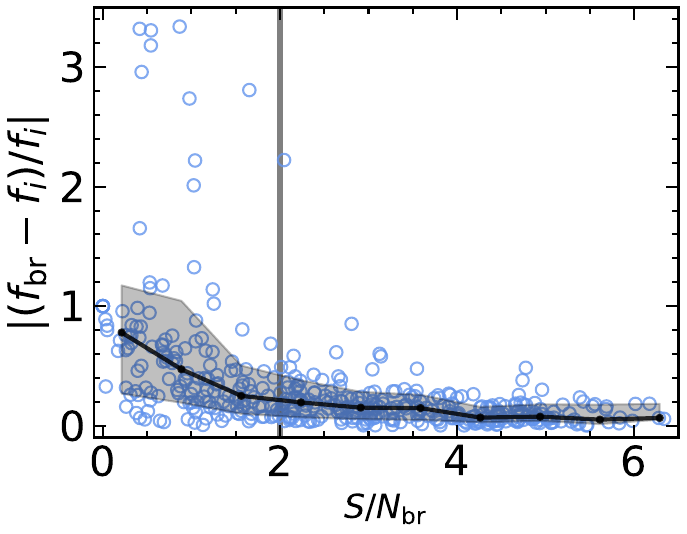}
\caption{Recovered broad-line fluxes from mock spectra. Error on the measured broad emission line flux $|(f_{\rm br} -f_i)/f_i|$, where $f_i$ is the input broad-line line flux and $f_{\rm br}$ is the measured broad-line flux, versus measured signal-to-noise ratio of simulated broad emission lines $S/N_{\rm br}$ with varying amplitudes on top of a noisy continuum. The black line and gray shaded region is the binned median and 1$\sigma$ uncertainty band. This demonstrates that the broad line detections and line flux measurements are very likely to be robust (to $\sim20$ percent) above $S/N_{\rm br} \approx 2$ (vertical gray line). }
\label{fig:snr}
\end{figure}

Given the significant uncertainty in our NIR spectra, we must determine whether the broad line detections are significant or not. We consider the broad emission line detection and associated flux measurement to be reliable using a signal-to-noise ratio criteria. We define the broad line signal-to-noise ratio as:
\begin{equation}
    S/N_{\rm br} = \frac{A_{\rm br}}{\rm MAD(resid)}
\end{equation}
where MAD is the median absolute deviation of the data minus our best-fit model residual (resid) in the window containing the line complex, and $A_{\rm br}$ is the peak or amplitude of the broad emission line. Exact definitions of the signal-to-noise ratio differ and depends on the details of the spectral modeling. We consider our broad line flux measurements to be reliable if $S/N_{\rm br} > 2$, which we justify using simulations. We simulate a single $\sim 4000$ km s$^{-1}$ broad Gaussian emission line (typical of our blazars) on top of a noisy power-law continuum, and repeat this procedure with varying emission line amplitudes. The recovered flux $f_{\rm br}$ is then compared to the true input flux $f_{i}$ by measuring the $S/N_{\rm br}$ from each simulated broad line. The $S/N_{\rm br} > 2$ threshold is determined as the approximate point where the error on the recovered flux is within $\sim20$, which is comparable to the systematic uncertainties from data reduction, flux calibration, and spectral modeling choices, as shown in Figure~\ref{fig:snr}.

\section{Results}\label{sec:results}

\subsection{Virial black hole mass estimates}

\begin{figure*}
\includegraphics[width=0.48\textwidth]{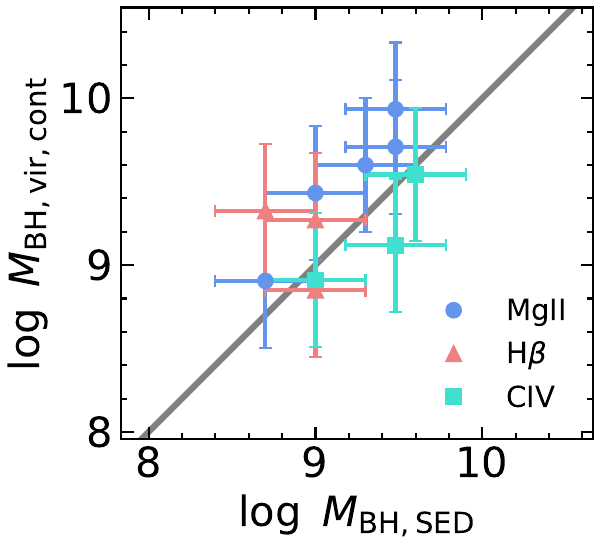}
\includegraphics[width=0.48\textwidth]{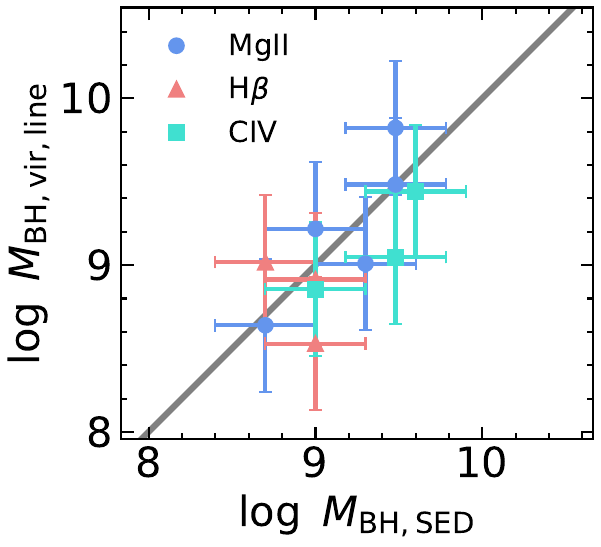}
\caption{Comparison between virial BH masses ($y$-axis) for the broad emission lines with $S/N>2$ in our GNIRS spectra and SED masses from \citet{Paliya2020} ($x$-axis) using the continuum (\emph{left}) or line luminosity (\emph{right}) and following the recipes in the text. The BH masses estimated from the continuum luminosity are slightly larger than those estimated from the broad line luminosity by $\sim 0.2$ dex. The gray $y=x$ line demonstrates there is no strong discrepancy between virial and SED -based BH mass estimates for our sample using either method. We assume typical uncertainties of 0.4 dex for the virial BH masses \citep{Shen2013} and 0.3 dex for the SED-based BH masses \citep{Paliya2020}.}
\label{fig:mass}
\end{figure*}

Following \citet{shen11}, we estimate the BH masses using the single-epoch virial method. This method assumes that the broad-line region (BLR) is virialized and uses the continuum luminosity and broad-line FWHM as a proxy for the BLR radius and virial velocity respectively. Under these assumptions, the BH mass can be estimated by:
\begin{equation}
\begin{split}
    \log{\left(\frac{M_{\rm{BH}} }{M_{\odot}} \right)} = a + b \log{ \left( \frac{L_{\rm{br}}}{10^{44}\ \rm{ erg\ s}^{-1}} \right) } + 2 \log{ \left( \frac{\rm{FWHM}_{\rm{br}}}{\rm{ km\ s}^{-1}} \right) }
\end{split}
\label{eq:BHmass}
\end{equation}
where $L_{\rm{br}}$ and FWHM$_{\rm{br}}$ are the broad-line luminosity and full-width-at-half-maximum (FWHM) with an intrinsic scatter of $\sim0.4$ dex in BH mass. The coefficients $a$ and $b$ are empirically calibrated against local AGNs with BH masses measured from reverberation mapping. We adopt the calibrations \citep{Vestergaard2006} used in \citet{shen11}:
\begin{equation}
    (a, b) = (0.910, 0.50), \quad \rm H\beta
\end{equation}
\vspace{-6mm}
\begin{equation}
    (a, b) = (0.660, 0.53), \quad \rm C\ IV
\end{equation}
\vspace{-6mm}
\begin{equation}
    (a, b) = (0.740, 0.62), \quad \rm Mg\ II.
\end{equation}

We caution that the continuum luminosities we have measured may be over-estimated if they are strongly relativistically beamed by the jet (e.g., \citealt{Wu2004,Shaw2012}), resulting in an over-estimation of the BH mass. Previous work has shown systematically-larger BH masses of 0.14 dex on average when broad line luminosities are used instead of continuum luminosities in the BH mass estimation prescriptions \citep{Shaw2012}. To estimate BH masses using the broad line luminosities, we make use of the fact that the line luminosity correlates with the continuum luminosity for broad-line quasars. We substitute the broad line luminosity into Equation~\ref{eq:BHmass} and adopt the calibrations derived by \citet{shen11} from \citet{Shaw2012}:
\begin{equation}
    (a, b) = (1.63, 0.49), \quad \rm H\beta
\end{equation}
\vspace{-6mm}
\begin{equation}
    (a, b) = (1.52, 0.46), \quad \rm C\ IV
\end{equation}
\vspace{-6mm}
\begin{equation}
    (a, b) = (1.70, 0.63), \quad \rm Mg\ II.
\end{equation}

We found that the masses derived from the continuum luminosities are $\sim 0.2$ dex larger than the masses derived from the broad line luminosities, roughly consistent with \citet{Shaw2012}. The BH mass comparisons are shown in Figure~\ref{fig:mass}.

\subsection{Eddington ratio estimates}

\begin{figure*}
\includegraphics[width=0.48\textwidth]{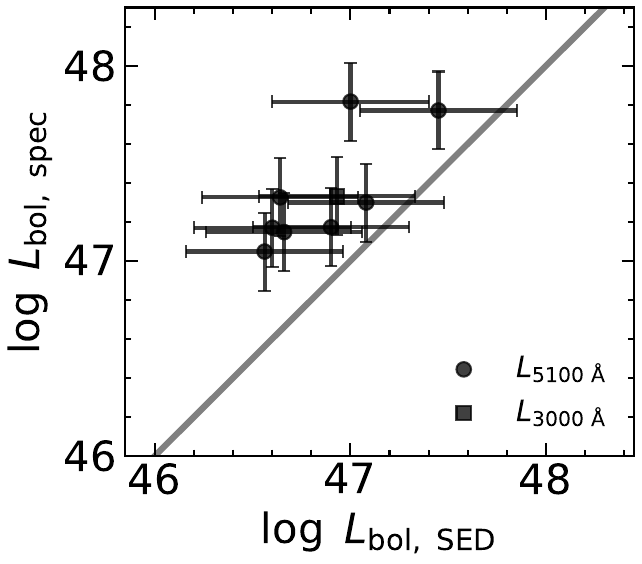}
\includegraphics[width=0.48\textwidth]{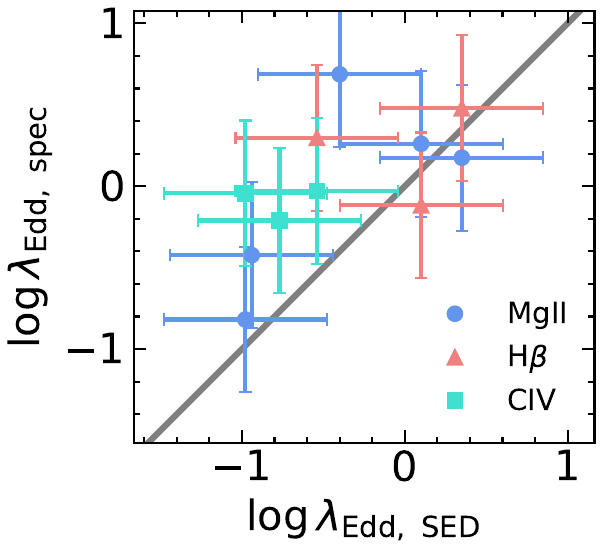}
\caption{\emph{Left:} Comparison between bolometric luminosities measured from the continuum of our NIR spectra using the 5100 \AA\ luminosity (or 3000 \AA\ luminosity for J1510+5702) ($y$-axis) and from the SED fitting approach ($x$-axis). The gray $y=x$ line demonstrates that the spectral continuum-based bolometric luminosities are, on average, larger than the SED fitting -based luminosities by $\sim 0.2$ dex. \emph{Right:} Comparison between Eddington luminosity ratios derived from the bolometric luminosity measured from the continuum and broad-line BH masses from the $M_{\rm{BH},\ vir,\ line}$ method from our NIR spectra ($y$-axis) and from the SED fitting approach ($x$-axis).  We assume typical uncertainties of 0.2 dex for spectral continuum -based bolometric luminosities ($L_{\rm bol,\ spec} = 9.26\ L_{\rm{5100}}$ or $5.15\ L_{\rm{3000}}$; \citealt{Richards2006,Runnoe2012}) and 0.4 dex for the SED fitting -based bolometric luminosities  ($L_{\rm bol,\ SED} = 2\ L_{\rm{disk}}$; \citealt{Paliya2020}). The gray $y=x$ line and the large uncertainties in the Eddington ratio estimates (summed-in-quadrature from the BH mass and bolometric luminosity uncertainties) demonstrates that they are broadly consistent with each other and that the Eddington ratios are $\lambda_{\rm Edd} \sim 0.1$--$1$. }
\label{fig:edd}
\end{figure*}

We estimate the Eddington ratios by diving the AGN bolometric luminosity by the Eddington luminosity ($\lambda_{\rm{Edd}} = L_{\rm{bol}} / L_{\rm{Edd}}$). Following e.g., \citet{Belladitta2022}, we estimate the AGN bolometric luminosity using two methods. First, following \cite{shen11}, we use the 5100 \AA\ (or 3000 \AA) continuum luminosities measured from our GNIRS spectra assuming bolometric corrections (BC) derived from a mean quasar SED \citep{Richards2006}. Those bolometric corrections are BC$_{5100} = 9.26$, BC$_{3000} = 5.15$. We ignore the small inclination factor of $i = \frac{\cos\ 0^\circ}{\cos\ 30^\circ} = 1.15$ in the bolometric correction expected to arise due to differences the viewing angle for Type 1 radio quiet AGNs and blazars. Secondly, we use the disk bolometric luminosities given by \cite{Paliya2020} inferred from SED fitting. The total AGN bolometric luminosity is computed by assuming $L_{\rm{bol}} = 2\ L_{\rm{disk}}$ (e.g., \citealt{Calderone2013}). To compute the Eddington luminosities, we usplote the SED-based BH masses against (denoted $\lambda_{\rm{Edd, SED}}$) our virial BH masses using the broad-line luminosites (denoted $\lambda_{\rm{Edd, spec}}$). The results are shown in Figure~\ref{fig:edd}. 

In Figure~\ref{fig:edd}, we show that the spectral continuum-based bolometric luminosities are, on average, larger than the SED fitting-based luminosities by $\sim 0.5$ dex (or about a factor of 3 in luminosity) for our sample. One possibility is that the systematic uncertainty in the bolometric correction ($\sim 0.2$ dex; \citealt{Runnoe2012}) is larger for blazars, or that the bolometric corrections for blazars are different. This could be explained by contamination of the quasar continuum emission  by a relativistic jet. Taken at face value, the discrepancy of about a factor of $\sim 3$ in luminosity is small given the Doppler factors of $\delta = 10 - 20$ of our parent sample of high-redshift FSRQs \citep{Paliya2020}. It is likely that only a portion of the UV/optical continuum emission is contaminated by beamed emission toward the observer, in contrast to the jetted radio emission, which is dominated by beamed emission (where Doppler factors are typically measured). This result demonstrates the importance of considering the caveats with virial BH masses and Eddington ratio calculations for blazars, which  may have implications for studies of how the Eddington luminosity ratios or accretion disk properties might relate to jet power (e.g., \citealt{Celotti2008}). 

\section{Summary and Conclusion}\label{sec:conclusion}

High-redshift Fermi blazars represent the most distant sources in the $\gamma$-ray sky, providing a laboratory to study early SMBH growth and super-Eddington accretion. They may form from the more rapid formation of jetted BHs, possibly by super-Eddington accretion, but their existing mass estimates are highly uncertain. These sources pose a challenge for models of early SMBH growth and formation. With Gemini/GNIRS spectroscopy of nine $z>3$ Fermi blazars, we have:
\begin{enumerate}
    \item Measured robust virial BH masses using broad \hbeta\ and/or \MgII\ for six of our nine target blazars with sufficient line signal-to-noise ratios. The \hbeta\ and \MgII\ virial masses have proven to be more robust than the existing \CIV -based masses. \CIV\ is known to be biased and prone to systematic uncertainties due to strong outflows \citep{Shen2012,Jun2015a}. 
    
    \item Compared the improved virial BH masses (with a 0.4 dex systematic scatter) against SED masses independently derived from modeling the accretion disk ($\sim$0.3 dex systematic uncertainty; \citealt{Paliya2020}). A significant discrepancy would imply deviation from the standard disk (e.g., due to high BH spins) in high-redshift blazars. We do not find a statistically significant discrepancy given the sample size. 

    \item Measured robust Eddington ratios and  bolometric luminosities from the \hbeta\ and/or \MgII\ broad lines and continuum luminosities for high-redshift blazars. We do find any compelling evidence to suggest they are super-Eddington accretors (e.g., \citealt{Begelman2017,Yang2020}).   

\end{enumerate} 

Future work with a larger sample size and could help confirm or falsify our main conclusions. This work highlights the challenge of obtaining deep enough NIR spectra to detect and model the broad emission lines sufficiently to estimate BH masses using ground-based observatories. Future observations with \emph{JWST} would be a compelling test for blazars at even higher redshifts.

\section*{Acknowledgements}

CJB and XL acknowledge support from NASA grant 80NSSC22K0030. YS acknowledges support from NSF grant AST-2009947. This research made use of \textsc{pypeit},\footnote{\url{https://pypeit.readthedocs.io/en/latest/}} a Python package for semi-automated reduction of astronomical slit-based spectroscopy \citep{pypeit:joss_pub, pypeit:zenodo}. This research made use of \textsc{Astroquery} \citep{Ginsburg2019}. This research made use of \textsc{Astropy}, a community-developed core Python package for Astronomy \citep{Astropy2018,Astropy2013}. We thank Gabriele Ghisellini, Tullia Sbarrato, and Jianfeng Wu for useful discussion. We are grateful to Lea Marcotulli for comments which improved our manuscript.

Based in part on observations obtained at the international Gemini Observatory (Program ID GN-2022A-Q-138; PI: X. Liu), a program of NSF’s NOIRLab, which is managed by the Association of Universities for Research in Astronomy (AURA) under a cooperative agreement with the National Science Foundation. on behalf of the Gemini Observatory partnership: the National Science Foundation (United States), National Research Council (Canada), Agencia Nacional de Investigaci\'{o}n y Desarrollo (Chile), Ministerio de Ciencia, Tecnolog\'{i}a e Innovaci\'{o}n (Argentina), Minist\'{e}rio da Ci\^{e}ncia, Tecnologia, Inova\c{c}\~{o}es e Comunica\c{c}\~{o}es (Brazil), and Korea Astronomy and Space Science Institute (Republic of Korea). This work was enabled by observations made from the Gemini North telescope, located within the Maunakea Science Reserve and adjacent to the summit of Maunakea. We are grateful for the privilege of observing the Universe from a place that is unique in both its astronomical quality and its cultural significance.

\section*{Data Availability}

The unreduced spectra will be available at the Gemini Observatory Archive at \url{https://archive.gemini.edu} after the proprietary period. The fully reduced, flux calibrated spectra will be made available upon reasonable request to the authors.




\bibliographystyle{mnras}
\bibliography{ref} 



\appendix

\section{NIR spectra}
\label{app:nir}

\begin{figure*}
\includegraphics[width=0.75\textwidth]{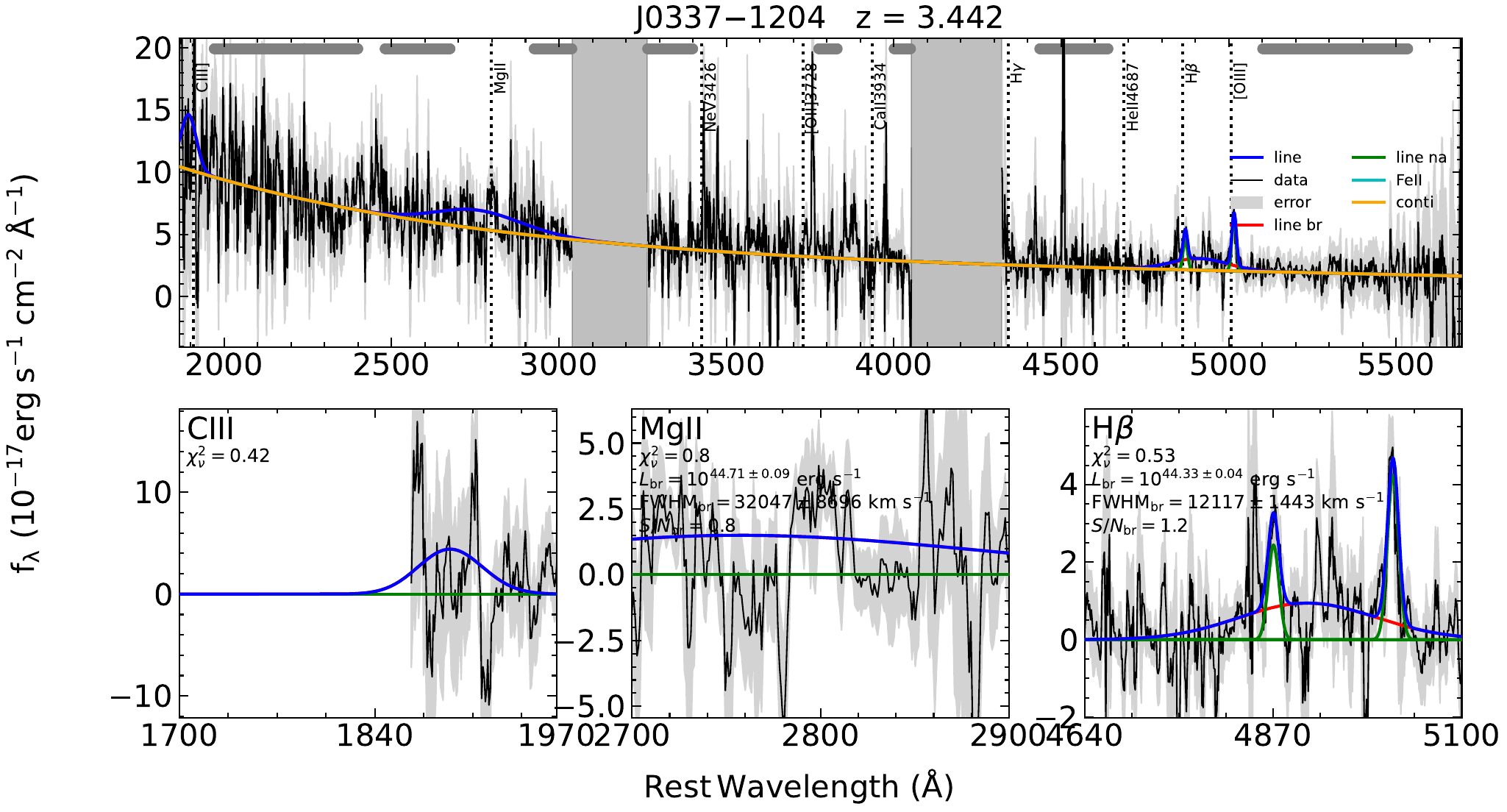}
\includegraphics[width=0.75\textwidth]{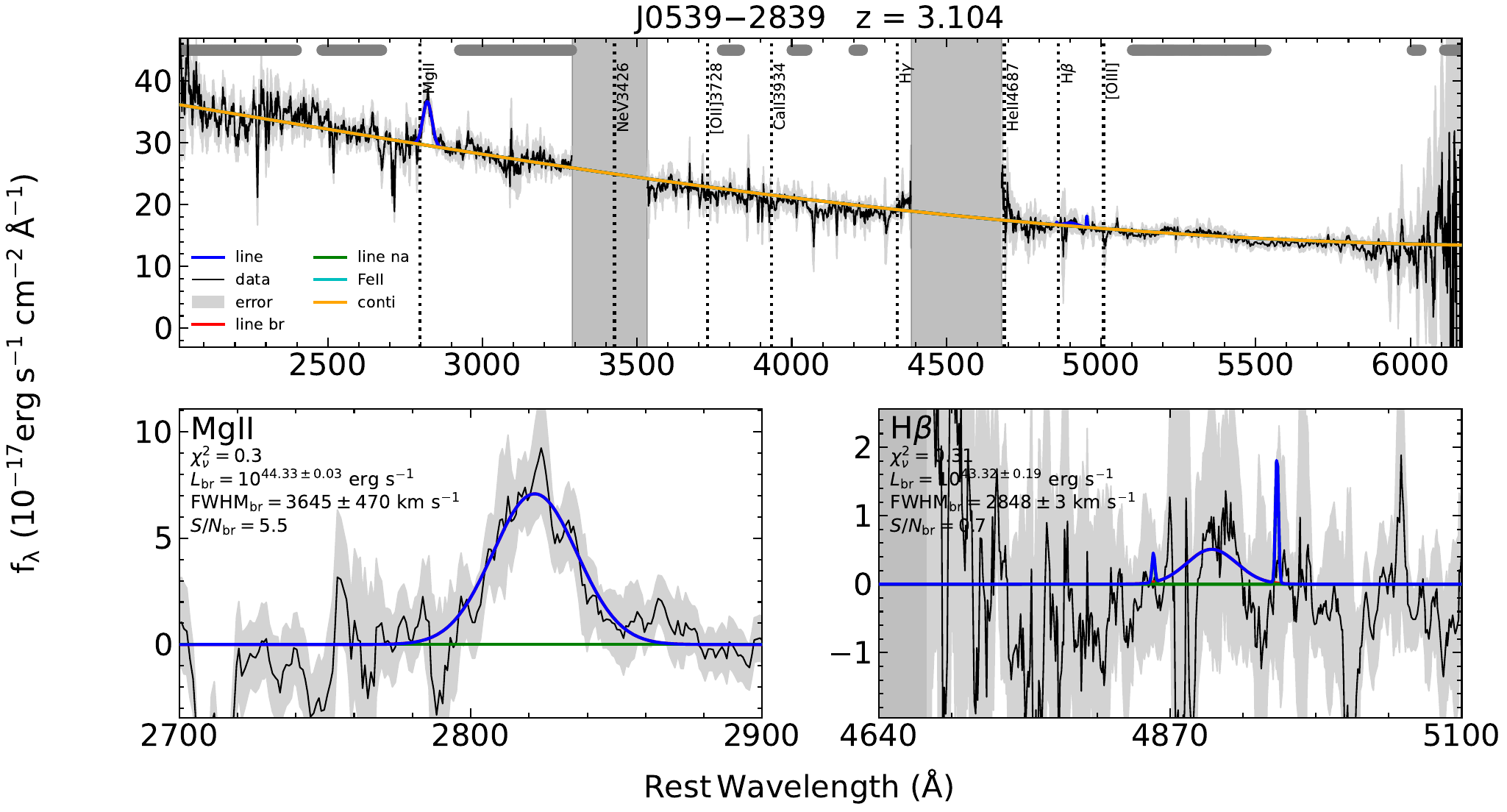}
\includegraphics[width=0.75\textwidth]{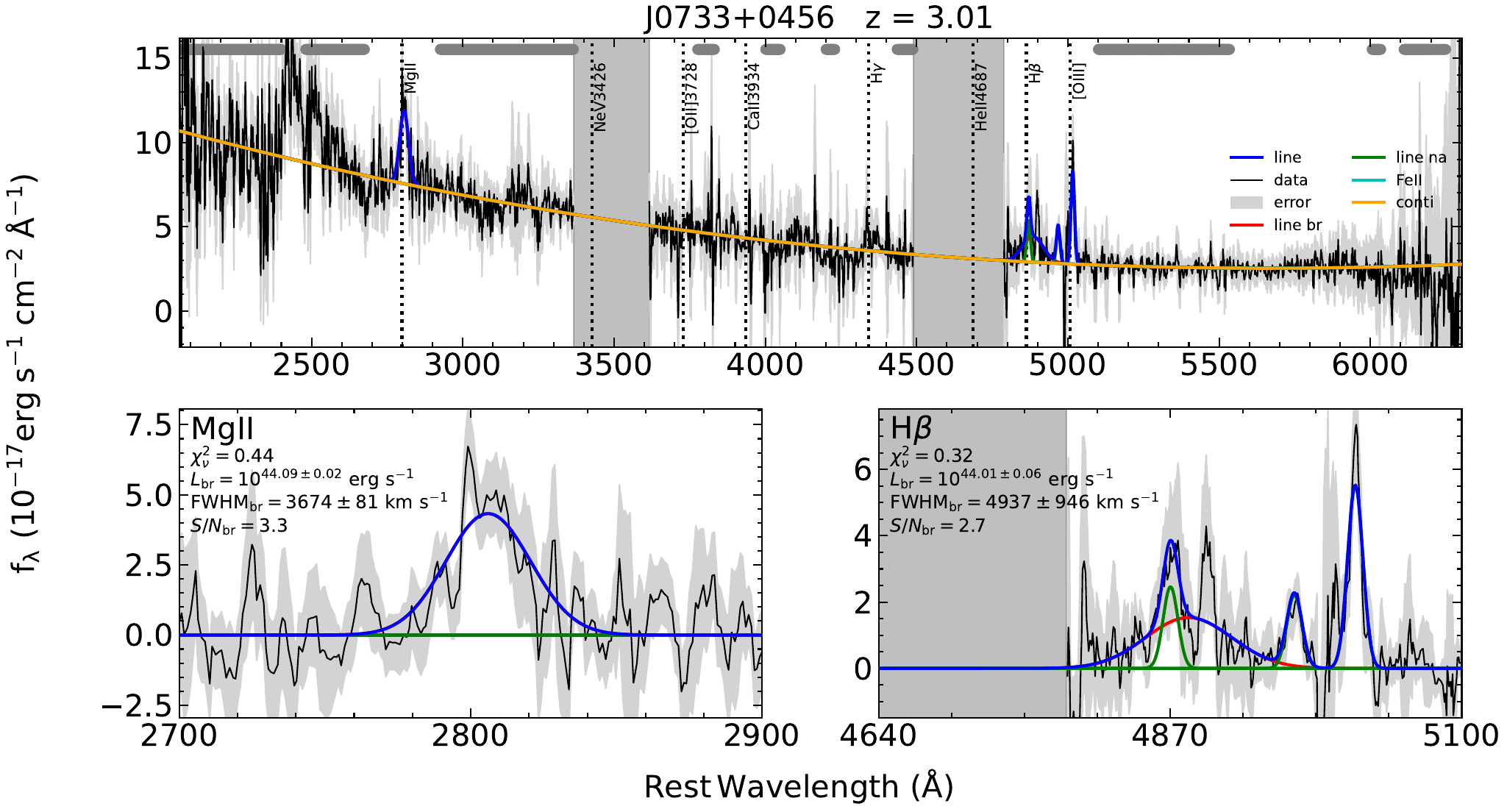}
\caption{GNIRS spectra of our sample. A power-law plus 3rd-order polynomial and Gaussians are used to fit the continuum (yellow) and emission lines, respectively. The Fe II emission templates (teal) are set to zero, as including them does not significantly improve the fits. The data is shown in black and the best-fit model is overplotted in blue. The individual narrow line components are plotted in green, and the broad line components are plotted in red. The gray bands on the top are line-free windows selected to determine the continuum emission. The light gray shaded bands are masked regions effected by telluric absorption at 1.35$-$1.45 $\mu$m and 1.8$-$1.92 $\mu$m. The lower panels show the zoomed-in emission line regions of \CIII , \MgII , and \hbeta\ as covered by the spectral wavelength range. The figure of the example source J0833-0454 is shown in Figure~\ref{fig:spec}. Only broad lines with $S/N_{\rm br}>2$ are considered reliable for measuring black hole masses.}
\label{fig:nir1}
\end{figure*}

\begin{figure*}
\includegraphics[width=0.75\textwidth]{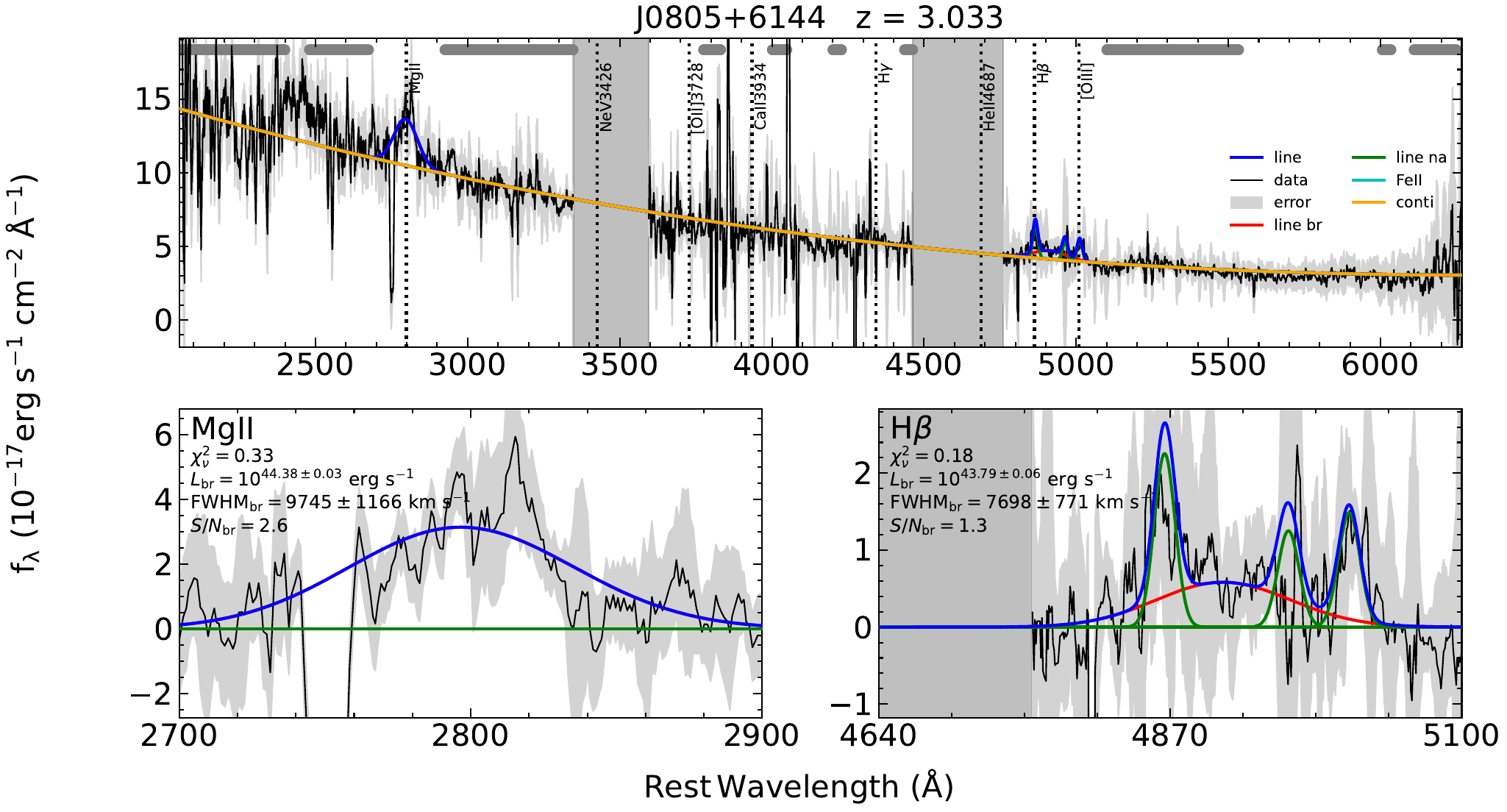}
\includegraphics[width=0.75\textwidth]{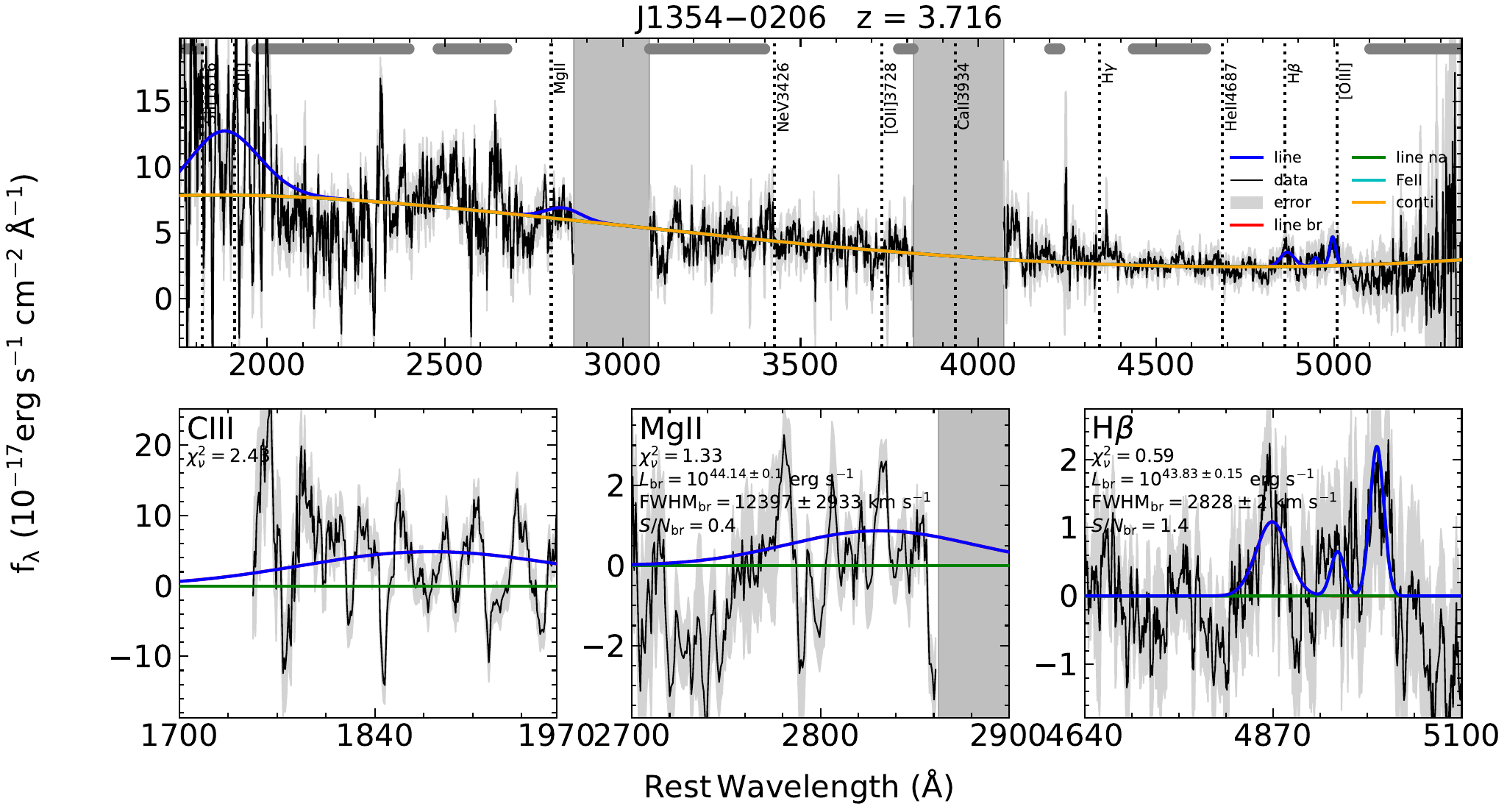}
\includegraphics[width=0.75\textwidth]{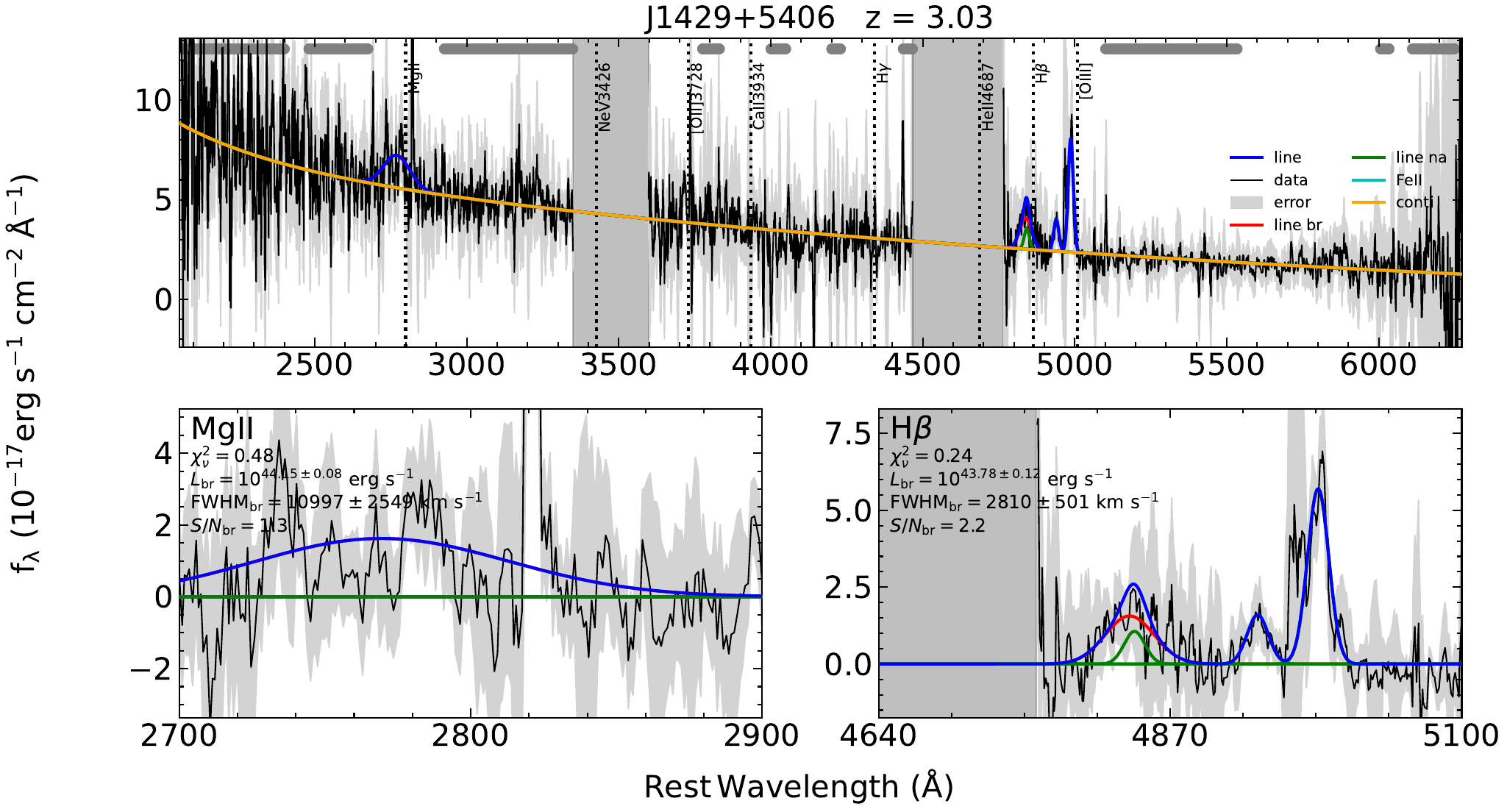}
\caption{Continued from Figure~\ref{fig:nir1}.}
\label{fig:nir2}
\end{figure*}

\begin{figure*}
\includegraphics[width=0.75\textwidth]{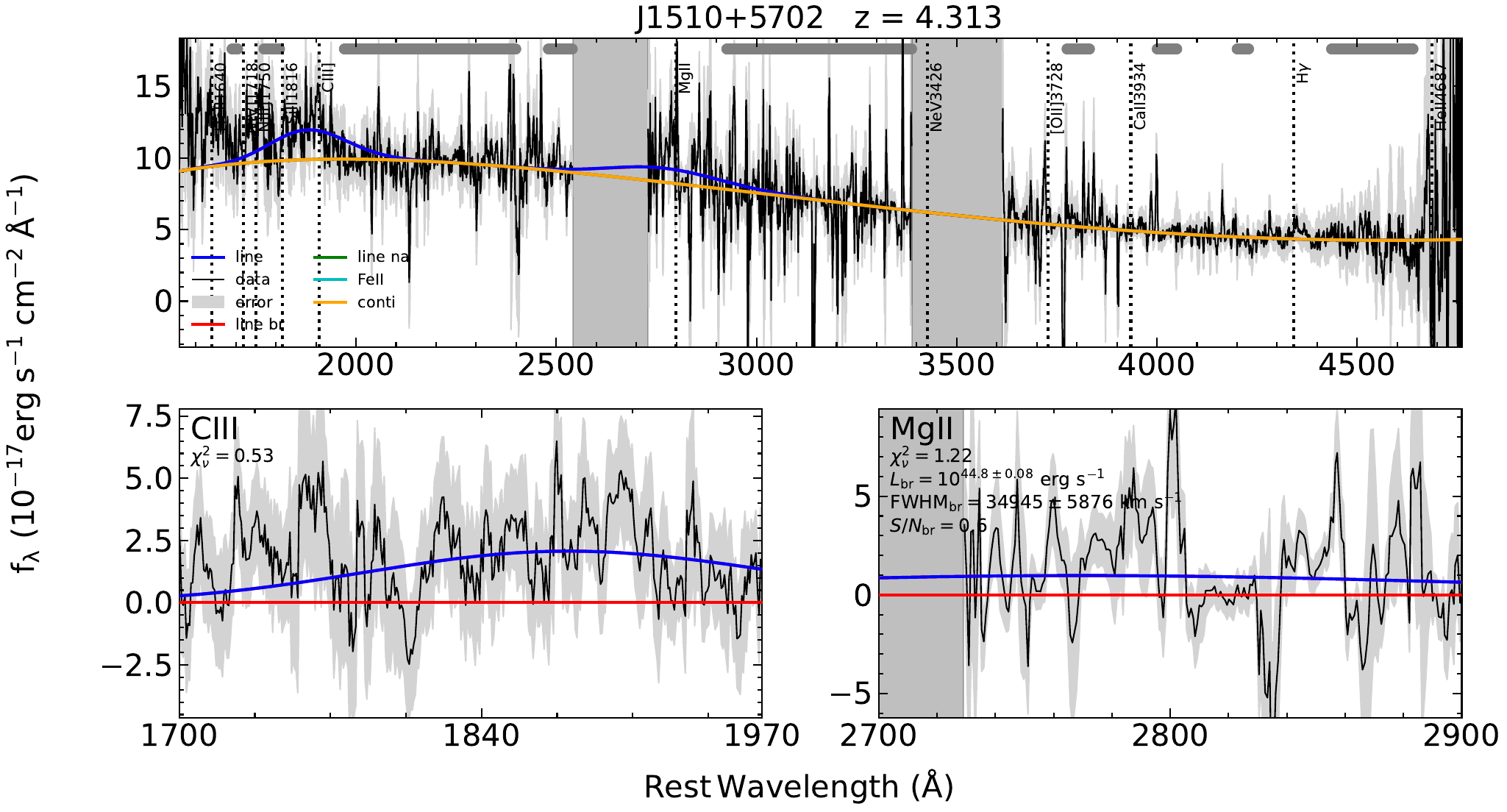}
\includegraphics[width=0.75\textwidth]{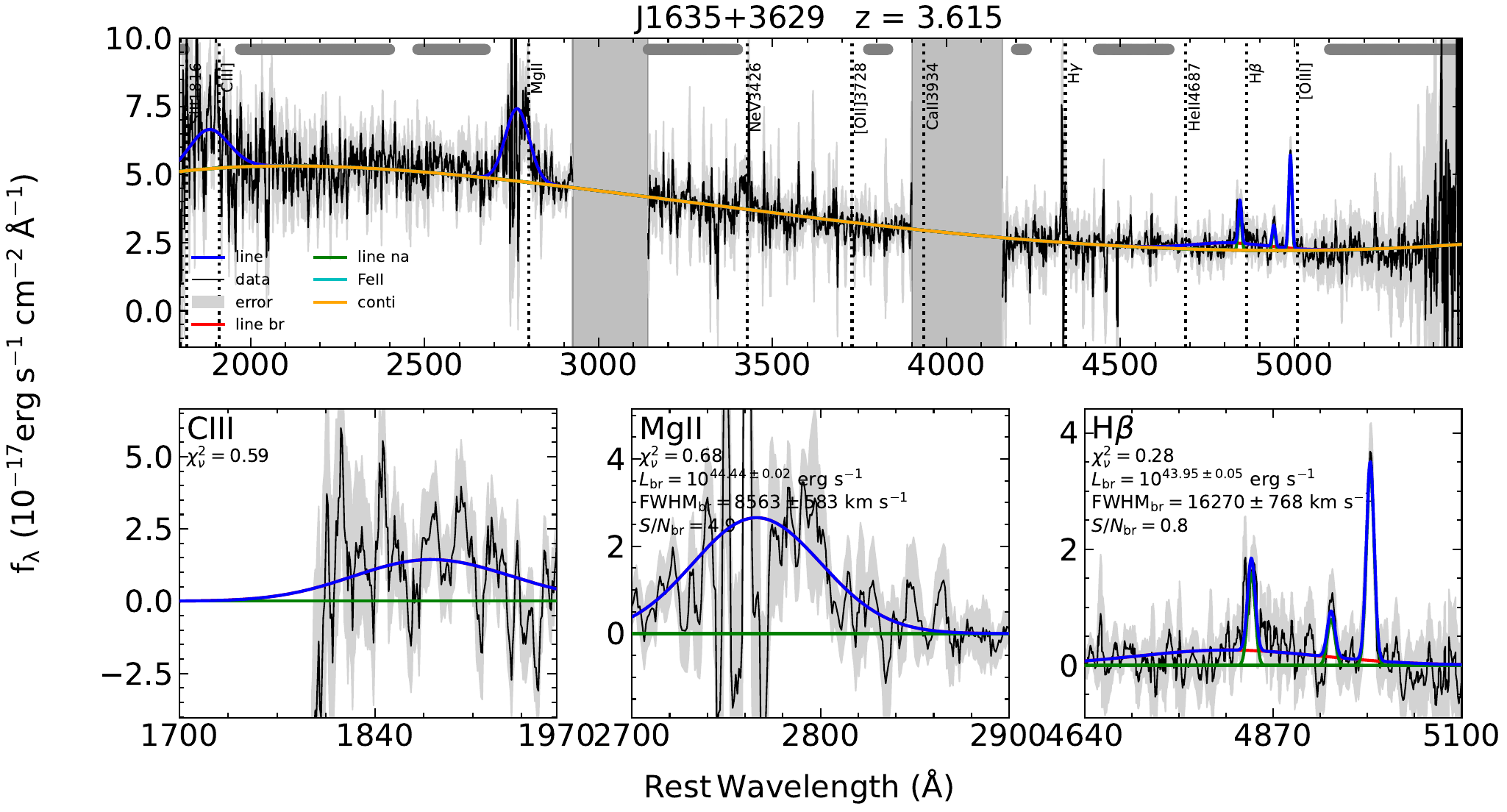}
\caption{Continued from Figure~\ref{fig:nir1}.}
\label{fig:nir3}
\end{figure*}

NIR spectra and modeling for all nine target blazars (Figures~\ref{fig:nir1}-\ref{fig:nir3}).





\bsp	
\label{lastpage}
\end{document}